\documentclass[journal]{IEEEtran}

\ifCLASSINFOpdf
\else
\fi
\usepackage{amsmath}
\allowdisplaybreaks
\usepackage{amssymb}
\usepackage{booktabs}
\usepackage{multirow}
\usepackage{graphicx}
\usepackage{algorithm}
\usepackage{algpseudocode}
\usepackage{tikz}
\usetikzlibrary{positioning, arrows.meta, calc}

\begin{document}

\title{Learning to Solve Two-Stage Stochastic Unit Commitment Problems with Quality Guarantees}

\author{Andrea~Fusco,
        Andrea~Lodi,
        and~Lavanya~Marla%
\thanks{This work has been submitted to the IEEE for possible publication. Copyright may be transferred without notice, after which this version may no longer be accessible.}%
\thanks{A.~Fusco is with Cornell Tech, New York, NY 10044 USA (e-mail: af655@cornell.edu).}%
\thanks{A.~Lodi is with the Jacobs Technion-Cornell Institute, Cornell Tech and Technion -- IIT, New York, NY 10044 USA (e-mail: al748@cornell.edu).}%
\thanks{L.~Marla is with the Department of Industrial and Enterprise Systems Engineering, University of Illinois Urbana-Champaign, Urbana, IL 61801 USA (e-mail: lavanyam@illinois.edu).}}

\markboth{Preprint -- submitted to the IEEE for possible publication}%
{Fusco \MakeLowercase{\textit{et al.}}: Learning to Solve Two-Stage Stochastic Unit Commitment Problems with Quality Guarantees}

\maketitle

\begin{abstract}
Two-stage stochastic Mixed-Integer Linear Programs are a canonical modeling tool to optimize power system operations under uncertainty, yet their extensive-form counterparts scale linearly with the number of scenarios and quickly become computationally prohibitive under day-ahead time constraints. We propose an Input Convex Neural Network architecture to learn a convex surrogate of the second-stage value function, enabling fast first-stage optimization while preserving convexity by construction. We couple the surrogate with a Neural-Benders correction loop that refines the first-stage solution a posteriori, recovering the exact optimum whether the network overestimates or underestimates the recourse cost, and certifying solution quality independently of surrogate accuracy. We evaluate the method on IEEE Stochastic Unit Commitment benchmarks (\texttt{case9}--\texttt{case118}) under both continuous and integer recourse. The proposed approach achieves solutions with zero optimality gap on the $K$-scenario instance across all test cases and scenario dimensions $K \in \{10, 50, 100\}$, with speedups up to $214\times$ over the Extensive Form in the integer setting. Solve times are stable across scenario realizations and compatible with day-ahead time windows on the tested benchmark instances.
\end{abstract}

\begin{IEEEkeywords}
Benders Decomposition, Input Convex Neural Networks, Optimality Guarantees, Surrogate Optimization, Two-stage Stochastic Programming, Unit Commitment.
\end{IEEEkeywords}

\IEEEpeerreviewmaketitle

\section*{Nomenclature}

\noindent\textit{Indices and Sets}
\begin{IEEEdescription}[\IEEEusemathlabelsep\IEEEsetlabelwidth{$\mathcal{E}^+(b),\mathcal{E}^-(b)$}]
\item[$g \in \mathcal{G}$] Thermal generators.
\item[$t \in \mathcal{T}$] Scheduling periods.
\item[$b \in \mathcal{B}$] Network buses.
\item[$e \in \mathcal{E}$] Transmission lines.
\item[$s \in \mathcal{S}$] Uncertainty scenarios.
\item[$\mathcal{G}_s,\, \mathcal{G}_f$] Slow and fast-start generator subsets ($\mathcal{G}_s \cup \mathcal{G}_f = \mathcal{G}$).
\item[$\mathcal{G}_b$] Generators connected to bus $b$.
\item[$\mathcal{E}^+(b),\mathcal{E}^-(b)$] Lines leaving and entering bus $b$.
\item[$i_e,\, j_e$] Sending and receiving buses of line $e$.
\end{IEEEdescription}

\noindent\textit{First-Stage Parameters}
\begin{IEEEdescription}[\IEEEusemathlabelsep\IEEEsetlabelwidth{$T^{up}_g,\, T^{dn}_g$}]
\item[$c^{su}_g$] Start-up cost of generator $g$ [\$].
\item[$c^{nl}_g$] No-load cost of generator $g$ [\$/h].
\item[$T^{up}_g,\, T^{dn}_g$] Minimum up- and down-time of generator $g$ [h].
\item[$p_f$] Marginal-cost percentile threshold for the slow/fast partition.
\end{IEEEdescription}

\noindent\textit{Second-Stage Parameters}
\begin{IEEEdescription}[\IEEEusemathlabelsep\IEEEsetlabelwidth{$\bar{p}_g,\, \underline{p}_g$}]
\item[$c_g$] Marginal generation cost of generator $g$ [\$/MWh].
\item[$c^{\mathrm{ls}}$] Value-of-lost-load penalty coefficient [\$/MWh].
\item[$\bar{p}_g,\, \underline{p}_g$] Maximum and minimum output of generator $g$ [MW].
\item[$r^u_g,\, r^d_g$] Ramp-up and ramp-down rates of generator $g$ [MW/h].
\item[$\beta_e$] Susceptance of line $e$ [p.u.].
\item[$\bar{f}_e$] Thermal limit of line $e$ [MW].
\item[$d_{b,t,s}$] Net load at bus $b$, period $t$, scenario $s$ [MW].
\end{IEEEdescription}

\noindent\textit{Decision Variables}
\begin{IEEEdescription}[\IEEEusemathlabelsep\IEEEsetlabelwidth{$(u^f\!,v^f\!,w^f)_{g,t,s}$}]
\item[$(u,v,w)_{g,t}$] Slow-generator commitment, start-up, shut-down binaries ($g \in \mathcal{G}_s$).
\item[$(u^f\!,v^f\!,w^f)_{g,t,s}$] Fast-generator commitment, start-up, shut-down binaries ($g \in \mathcal{G}_f$).
\item[$p_{g,t,s}$] Power output of generator $g$, period $t$, scenario $s$ [MW].
\item[$\ell_{b,t,s}$] Load shedding at bus $b$, period $t$, scenario $s$ [MW].
\item[$f_{e,t,s}$] Power flow on line $e$, period $t$, scenario $s$ [MW].
\item[$\theta_{b,t,s}$] Voltage angle at bus $b$, period $t$, scenario $s$ [rad].
\end{IEEEdescription}

\noindent\textit{Functions}
\begin{IEEEdescription}[\IEEEusemathlabelsep\IEEEsetlabelwidth{$\mathbb{E}_\xi[Q(x,\xi)]$}]
\item[{$Q(x,\xi_s)$}] Second-stage recourse value function under scenario $s$.
\item[{$\mathbb{E}_\xi[Q(x,\xi)]$}] Expected recourse cost.
\item[{$\hat{Q}(x,z)$}] Input Convex Neural Network surrogate of the recourse cost.
\end{IEEEdescription}

\noindent\textit{Scenario Encoding and Neural Surrogate}
\begin{IEEEdescription}[\IEEEusemathlabelsep\IEEEsetlabelwidth{$W_x^{(l)},W_z^{(l)},W_h^{(l)}$}]
\item[$K$] Number of sampled scenarios.
\item[$p_k$] Probability of scenario $k$.
\item[$\Psi_1$] Per-scenario encoder network.
\item[$\Psi_2$] Scenario aggregation network.
\item[$n_x$] Dimension of the first-stage decision vector $x$.
\item[$d_2$] Dimension of the aggregated scenario embedding $z$.
\item[$z \in \mathbb{R}^{d_2}$] Aggregated scenario embedding (computed offline).
\item[$L$] Number of hidden layers in the ICNN.
\item[$h^{(l)}$] Hidden-layer activation vector at layer $l$, $l=1,\ldots,L$.
\item[$W_x^{(l)},W_z^{(l)},W_h^{(l)}$] ICNN weight matrices at layer $l$ (input, scenario, passthrough).
\item[$\sigma(\cdot)$] ReLU activation: $\sigma(t)=\max\{0,t\}$.
\item[$\delta_j^{(l)}$] Binary activation indicator of neuron $j$ at layer $l$ (MILP reformulation).
\end{IEEEdescription}

\noindent\textit{Neural-Benders Correction Loop}
\begin{IEEEdescription}[\IEEEusemathlabelsep\IEEEsetlabelwidth{$\overline{Q}(x^{(i)})$}]
\item[$\rho$] Slack penalty parameter.
\item[$\lambda$] Slack on the ICNN constraint of the master.
\item[$\eta$] Auxiliary recourse cost estimator.
\item[$\theta(x)$] Cut-based lower bound on the recourse cost.
\item[$h_k,\,T_k$] Right-hand side $h(\xi_k)$ and technology matrix $T(\xi_k)$ of scenario $k$.
\item[$\pi_k^*$] LP duals of scenario $k$ (continuous recourse).
\item[$d_k^*$] Dual unbounded ray of scenario $k$ (Farkas cut).
\item[$L$] LP-relaxation lower bound on the recourse cost.
\item[$S^+,\,S^-$] Index sets where $u^{(i)}_{g,t}=1$ and $u^{(i)}_{g,t}=0$.
\item[$\overline{Q}(x^{(i)})$] Sample upper bound on the recourse cost at $x^{(i)}$.
\item[$\mathrm{gap}^{(i)}$] Solution quality gap at iteration $i$.
\item[$\varepsilon$] Convergence tolerance.
\end{IEEEdescription}

\section{Introduction}
\label{sec:intro}
\IEEEPARstart{T}{he} large-scale integration of renewable energy sources (primarily wind and solar photovoltaic) introduces significant uncertainty into the generation mix, as output forecasts can deviate substantially from actual realizations. In this context, the classical deterministic Unit Commitment (UC) model, which schedules generating units under the assumption of perfectly known demand and generation, can yield infeasible or highly suboptimal solutions, motivating the adoption of stochastic optimization frameworks.

The Stochastic Unit Commitment (SUC) problem addresses this challenge by optimizing commitment schedules over a distribution of possible uncertainty realizations. SUC problems are, however, notoriously difficult. As stochastic mixed-integer programs, they are NP-hard, and their tractability degrades rapidly with the dimensionality of the uncertainty. Accurately representing high-dimensional renewable variability requires a large number of scenarios, yielding Mixed-Integer Linear Programs (MILPs) of rapidly growing size that can quickly exceed the capabilities of commercial solvers within the tight time budgets imposed by day-ahead market clearing.

\subsection{Literature Review}
\label{sec:litreview}
The integration of large-scale renewable energy has made two-stage stochastic programming a widely adopted modeling framework for the SUC problem \cite{Birge2011}.
Under the Sample Average Approximation (SAA) \cite{Kleywegt2002}, the expected recourse term is approximated by an empirical average over $K$ sampled realizations of the uncertainty, resulting in a finite-scenario stochastic program.
The corresponding extensive-form (deterministic-equivalent) MILP replicates the second-stage variables and constraints for each scenario (while enforcing first-stage non-anticipativity), and its size therefore grows linearly with $K$.

Classical exact methods address this scalability challenge by exploiting the block-angular structure of the deterministic equivalent.
Benders Decomposition (known in the stochastic programming literature as the L-Shaped Method) iterates between a master problem that handles the binary commitment decisions and $K$ independent scenario subproblems, progressively refining the expected recourse cost approximation via dual optimality and feasibility cuts \cite{Papavasiliou2015}.
While theoretically rigorous when the recourse set $Y$ is continuous, as in standard Direct Current (DC) economic dispatch, these algorithms suffer from two fundamental bottlenecks.
Their per-iteration cost scales linearly with $K$, since a full set of subproblems must be solved at each step.
More critically, when recourse involves integer variables (such as fast-start unit activation or transmission line switching) the value function $Q(x, \xi)$ loses convexity and piecewise linearity with respect to $x$, invalidating the strong duality arguments upon which Benders cuts rely.
Extensions such as the Integer L-Shaped Method and Stochastic Dual Dynamic Integer Programming recover correctness in this setting but introduce substantial additional algorithmic complexity \cite{Laporte1993, Caroe1998, Zou2019}.

Scenario reduction methods mitigate this bottleneck by selecting a smaller subset that approximates the original distribution in probability distance, either by iteratively discarding or aggregating scenarios \cite{Dupacova2003} or by learning compact embeddings tailored to the downstream optimization task \cite{ScenarioReduction2024}.
Either way, the reduced distribution carries an approximation error, and no mechanism bounds the resulting suboptimality of the first-stage solution.

In response to the computational intractability of exact solvers, recent research has pivoted towards machine learning \cite{Yang2021}.
One prominent direction uses offline predictors to prune or warm-start the MILP: reducing the binary variable space \cite{Zhou2021, Lin2022}, or providing high-quality initial solutions \cite{Xavier2021}.
End-to-end deep learning and deep reinforcement learning attempt to bypass the solver entirely by mapping system parameters to dispatch decisions \cite{Donti2017}.
Although these approaches achieve near-instantaneous inference, they inherently struggle to enforce hard physical constraints, yielding black-box solutions that may violate power balance, line limits, or inter-temporal generator requirements \cite{Chen2021}.
To bridge computational speed with operational feasibility, the Neural Two-Stage Stochastic Programming (Neur2SP) framework and its variants embed a trained neural surrogate of the recourse cost into the first-stage MILP, delegating constraint enforcement to the exact solver \cite{Dumouchelle2022, Kronqvist2023, Alcantara2024}.
Advances in MILP formulations for ReLU-activated networks \cite{Fischetti2018, Anderson2020} have made this embedding tractable.
However, since Neur2SP relies on standard multi-layer perceptrons (MLPs), the resulting cost approximation is generally non-convex with respect to the first-stage decision variables, which can introduce spurious basins into the surrogate optimization landscape and prevents the network itself from serving as a valid lower bound on the recourse cost.
More critically, Neur2SP is a one-shot method. It solves the surrogate MILP once and returns a solution with no refinement loop, so no a posteriori certificate of solution quality is available. This certification gap is what the present work closes.

\subsection{Contribution}
We propose a learning framework for two-stage SUC that extends the Neur2SP paradigm with two modifications: an Input Convex Neural Network (ICNN) surrogate and a hybrid Neural-Benders refinement loop. The main contributions can be summarized as follows:
\begin{itemize}
\item We adopt an ICNN surrogate for the recourse cost, choosing this architecture for three structural reasons: in the continuous setting, $\mathbb{E}_\xi[Q(x,\xi)]$ is convex and piecewise linear in $x$, so the ICNN approximates the target with a model of the same class; convexity by construction eliminates spurious basins from the first-stage optimization that a non-convex surrogate could otherwise introduce; and the convex continuous relaxation of the embedded MILP yields a tighter LP bound per master iteration.
\item We develop a hybrid Neural-Benders correction loop that certifies and tightens the first-stage solution a posteriori; the optimality certificate rests on accumulated exact-subproblem cuts and scenario evaluation, independent of surrogate accuracy, closing the certification gap that Neur2SP cannot address. The ICNN warm start eliminates spurious basins, so the loop converges in 1--13 iterations across the benchmark against up to 100 uncertified iterations for the classical Integer L-Shaped method.
\item We extend the methodology to the integer recourse setting (intractable for classical decompositions) by leveraging the convex structure of the ICNN surrogate, yielding certified first-stage schedules even in the presence of discrete second-stage decisions.
\end{itemize}

Section~\ref{sec:background} introduces the two-stage stochastic programming formulation and the SUC problem. Section~\ref{sec:methodology} details the proposed methodology. Section~\ref{sec:experiments} describes the experimental settings. Section~\ref{sec:results} presents the numerical results. Section~\ref{sec:conclusion} concludes the paper.

\section{Background}
\label{sec:background}
\subsection{Two-Stage Stochastic Programming}
\label{sec:twostage}
Let $x \in X \subseteq \mathbb{R}^{n_1}$ denote the first-stage decision variables, where $X$ is defined by a set of deterministic constraints, and let $\xi \in \Xi$ be a random vector representing the uncertain parameters. Once $\xi$ is realized, second-stage recourse variables $y \in \mathbb{R}^{n_2}$ are chosen to satisfy operational requirements. The general two-stage stochastic program is formulated as
\begin{equation}
    \min_{x \in X} \left\{ c^\top x + \mathbb{E}_{\xi}[Q(x, \xi)] \right\},
\end{equation}
where $Q(x, \xi)$ is the optimal value of the second-stage recourse problem
\begin{equation}
    Q(x, \xi) = \min_{y} \left\{ q^\top y \mid Wy = h(\xi) - Tx, \; y \in Y \right\}.
\end{equation}
Here, $c$ and $q$ are cost vectors, $T$ is the technology matrix linking first- and second-stage decisions, $W$ is the recourse matrix, and $h(\xi)$ captures the dependence of the right-hand side on the uncertain parameters.

When the recourse set $Y$ is continuous, $Q(x, \xi)$ is convex and piecewise linear in $x$ for each fixed $\xi$ \cite{Birge2011}, so $\mathbb{E}_{\xi}[Q(x, \xi)]$ is convex in $x$. This property underlies the validity of Benders Decomposition and the L-Shaped Method, which approximate the recourse function through supporting hyperplanes derived from dual information, and is the structural property the ICNN surrogate of Section~\ref{sec:icnn} is designed to match. When integer variables are present in the recourse, $Q(x, \xi)$ is generally non-convex and discontinuous in $x$, invalidating dual-based cuts and rendering classical decomposition inapplicable without substantial extensions.

\subsection{Stochastic Unit Commitment Problem}
\label{sec:suc}
The SUC problem is a natural instance of the two-stage stochastic framework introduced above. Let $\mathcal{G}$, $\mathcal{T}$, $\mathcal{B}$, and $\mathcal{E}$ denote the sets of thermal generators, scheduling periods, buses, and transmission lines, respectively. The uncertain parameter $\xi$ is indexed by scenarios $s \in \mathcal{S}$ and captures the net load $d_{b,t,s}$ at each bus $b \in \mathcal{B}$ and period $t \in \mathcal{T}$, aggregating load demand and renewable generation variability.

\subsubsection{First Stage (Unit Commitment)}
The first-stage decision $x = (u, v, w)$ collects binary variables: $u_{g,t} \in \{0,1\}$ is the on/off status of generator $g \in \mathcal{G}$ at period $t \in \mathcal{T}$, while $v_{g,t}$ and $w_{g,t}$ are start-up and shut-down indicators. The first-stage cost is $c^{\top}x = \sum_{g,t}(c^{su}_g v_{g,t} + c^{nl}_g u_{g,t})$, where $c^{su}_g$ and $c^{nl}_g$ are start-up and no-load costs, respectively. The feasible set $X$ is defined by the following constraints, for all $g \in \mathcal{G}$, $t \in \mathcal{T}$
\begin{subequations}
\begin{align}
    v_{g,t} - w_{g,t} &= u_{g,t} - u_{g,t-1} \label{eq:logical} \\
    \sum_{\tau=t-T^{up}_g+1}^{t} v_{g,\tau} &\leq u_{g,t} \label{eq:minup} \\
    \sum_{\tau=t-T^{dn}_g+1}^{t} w_{g,\tau} &\leq 1 - u_{g,t}, \label{eq:mindown}
\end{align}
\end{subequations}
where $T^{up}_g$ and $T^{dn}_g$ are minimum up- and down-time requirements. Constraint~\eqref{eq:logical} links commitment, start-up, and shut-down decisions, while \eqref{eq:minup}--\eqref{eq:mindown} enforce minimum operating durations.

\subsubsection{Second Stage (Economic Dispatch)}
Given a commitment schedule $x$ and a scenario $s \in \mathcal{S}$, the second-stage problem determines the least-cost dispatch. Let $p_{g,t,s} \geq 0$ be the output of generator $g$, $f_{e,t,s}$ the power flow on line $e \in \mathcal{E}$, and $\theta_{b,t,s}$ the voltage angle at bus $b$. The recourse value function is
\begin{equation}
    Q(x, \xi_s) = \min_{p_s,\, f_s,\, \theta_s,\, \ell_s} \;\; \sum_{g \in \mathcal{G}} \sum_{t \in \mathcal{T}} c_g \, p_{g,t,s} \;+\; c^{\mathrm{ls}} \sum_{b \in \mathcal{B}} \sum_{t \in \mathcal{T}} \ell_{b,t,s},
\end{equation}
subject to, for all $b \in \mathcal{B}$, $e \in \mathcal{E}$, $g \in \mathcal{G}$, $t \in \mathcal{T}$
\begin{subequations}
\begin{align}
    \sum_{g \in \mathcal{G}_b} p_{g,t,s} + \ell_{b,t,s} - d_{b,t,s} &= \sum_{e \in \mathcal{E}^+(b)} \!\! f_{e,t,s} - \sum_{e \in \mathcal{E}^-(b)} \!\! f_{e,t,s} \label{eq:balance} \\
    f_{e,t,s} &= \beta_e \bigl(\theta_{i_e,t,s} - \theta_{j_e,t,s}\bigr) \label{eq:dcflow} \\
    -\bar{f}_e &\leq f_{e,t,s} \leq \bar{f}_e \label{eq:linelimit} \\
    \underline{p}_g\, u_{g,t} &\leq p_{g,t,s} \leq \bar{p}_g\, u_{g,t} \label{eq:genbounds} \\
    p_{g,t,s} - p_{g,t-1,s} &\leq r^u_g \label{eq:rampup} \\
    p_{g,t-1,s} - p_{g,t,s} &\leq r^d_g, \label{eq:rampdown}
\end{align}
\end{subequations}
where $\mathcal{G}_b$ is the set of generators at bus $b$; $\mathcal{E}^+(b)$ and $\mathcal{E}^-(b)$ are the sets of lines leaving and entering bus $b$; $\beta_e$ and $\bar{f}_e$ are the susceptance and thermal limit of line $e$; $\bar{p}_g$, $\underline{p}_g$ are output bounds; and $r^u_g$, $r^d_g$ are ramp-up and ramp-down rates. The variable $\ell_{b,t,s} \geq 0$ represents load shedding at bus $b$, period $t$, scenario $s$, penalized by the value-of-lost-load coefficient $c^{\mathrm{ls}}$, which is set orders of magnitude larger than any marginal generation cost. Constraint~\eqref{eq:balance} enforces nodal power balance via the DC power flow model~\eqref{eq:dcflow}; \eqref{eq:genbounds} conditions generation on the commitment state; and \eqref{eq:rampup}--\eqref{eq:rampdown} limit inter-temporal output variations.

In the continuous recourse setting, $y_s = (p_s, f_s, \theta_s) \in \mathbb{R}_+^{|\mathcal{G}||\mathcal{T}|} \times \mathbb{R}^{|\mathcal{E}||\mathcal{T}|} \times \mathbb{R}^{|\mathcal{B}||\mathcal{T}|}$ and $Q(x, \xi_s)$ is convex in $x$ for each fixed scenario $s$. In the integer recourse variant, the second stage additionally includes binary activation of fast-start units or transmission switching decisions, rendering $Q(x, \xi_s)$ non-convex and invalidating standard decomposition methods.

\subsection{SUC with Integer Recourse: Fast-Start Unit Commitment}
\label{sec:strategic}
We extend the IEEE benchmark SUC of Section~\ref{sec:suc} to incorporate \emph{integer recourse} by distinguishing between two classes of generating units: slow generators, whose commitment must be decided before uncertainty is revealed, and fast-start generators, which can be activated on short notice after the load realization is observed.

The generator set $\mathcal{G}$ is partitioned into slow generators $\mathcal{G}_s$ and fast-start generators $\mathcal{G}_f = \mathcal{G} \setminus \mathcal{G}_s$. The partition is defined by marginal cost rank: $\mathcal{G}_f$ comprises the $\lceil (1 - p_f)\,|\mathcal{G}| \rceil$ generators with the highest marginal costs $c_g$, for a percentile parameter $p_f \in (0,1)$, reflecting the merit-order rationale that expensive peakers are the natural candidates for flexible, post-uncertainty activation.

\subsubsection{First Stage}
The first-stage decision $x = (u, v, w)$ governs the slow generators $g \in \mathcal{G}_s$ only. Commitment logic~\eqref{eq:logical} and minimum operating-time constraints~\eqref{eq:minup}--\eqref{eq:mindown} apply exclusively to $\mathcal{G}_s$, and the first-stage cost is $c^\top x = \sum_{g \in \mathcal{G}_s,\, t} (c^{su}_g\, v_{g,t} + c^{nl}_g\, u_{g,t})$.

\subsubsection{Second Stage}
Given first-stage decisions $x$ and a scenario realization $\xi_s = d_{\cdot,\cdot,s}$, the second stage solves a \emph{mixed-integer} dispatch problem over both generator classes. For each fast generator $g \in \mathcal{G}_f$ and scenario $s$, the wait-and-see binary variables $u^f_{g,t,s},\, v^f_{g,t,s},\, w^f_{g,t,s} \in \{0,1\}$ govern scenario-specific commitment decisions, namely
\begin{subequations}
\label{eq:fast-logic}
\begin{align}
    v^f_{g,t,s} - w^f_{g,t,s} &= u^f_{g,t,s} - u^f_{g,t-1,s}, \label{eq:fast-logic-a} \\
    \textstyle\sum_{\tau=t-T^{up}_g+1}^{t} v^f_{g,\tau,s} &\leq u^f_{g,t,s}, \label{eq:fast-minup} \\
    \textstyle\sum_{\tau=t-T^{dn}_g+1}^{t} w^f_{g,\tau,s} &\leq 1 - u^f_{g,t,s}. \label{eq:fast-mindown}
\end{align}
\end{subequations}
The recourse value function is
\begin{equation}
\label{eq:q-integer}
Q(x, \xi_s) = \min_{\substack{p_s,\, f_s,\, \theta_s \\ u^f_s,\, v^f_s,\, w^f_s}} \;\; \sum_{g \in \mathcal{G}} \sum_{t \in \mathcal{T}} c_g\, p_{g,t,s},
\end{equation}
subject to the power balance~\eqref{eq:balance}, DC flow~\eqref{eq:dcflow}--\eqref{eq:linelimit}, and ramping constraints~\eqref{eq:rampup}--\eqref{eq:rampdown} for all $g \in \mathcal{G}$, together with \eqref{eq:fast-logic} for $g \in \mathcal{G}_f$, and the generation bounds
\begin{equation}
\label{eq:gen-bounds-split}
\underline{p}_g\, \bar{u}_{g,t,s} \leq p_{g,t,s} \leq \bar{p}_g\, \bar{u}_{g,t,s}, \quad
\bar{u}_{g,t,s} := \begin{cases} u_{g,t} & g \in \mathcal{G}_s, \\ u^f_{g,t,s} & g \in \mathcal{G}_f. \end{cases}
\end{equation}

Load-shedding variables $\ell_{b,t,s} \geq 0$ ensure that the nodal balance constraints can always be satisfied, endowing the formulation with complete recourse by construction.

The binary variables $\{u^f_{g,t,s}\}_{g \in \mathcal{G}_f}$ in the second stage render $Q(x,\xi_s)$ non-convex and generally discontinuous in $x$, as observed in Section~\ref{sec:twostage}. This invalidates the Linear Programming (LP) duality-based cuts of the standard L-Shaped Method, which are valid lower bounds only when the recourse is continuous. In the Neural-Benders correction loop of Section~\ref{sec:benders}, correctness for this setting is recovered by replacing LP optimality cuts with combinatorial cuts derived from the integer subproblem evaluated at each candidate $x^{(i)}$~\cite{Laporte1993, Zou2019}.

\section{Methodology}
\label{sec:methodology}
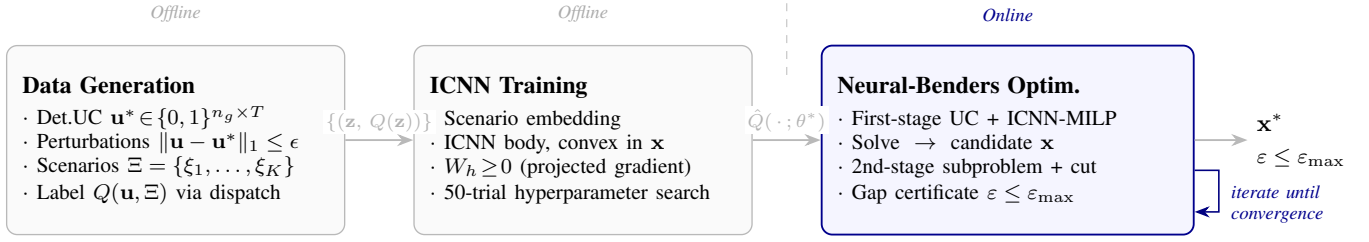
\begin{figure*}[!t]
\centering
\begin{tikzpicture}[
  font=\small,
  >=Stealth,
  offbox/.style={
    draw=gray!55, rounded corners=4pt, fill=gray!4,
    text width=4.05cm, inner sep=6pt, align=left, line width=0.5pt,
    minimum height=2.5cm,
  },
  onbox/.style={
    draw=blue!55!black, rounded corners=4pt, fill=blue!4,
    text width=4.5cm, inner sep=6pt, align=left, line width=0.7pt,
    minimum height=2.5cm,
  },
  arr/.style={-Stealth, draw=gray!55, line width=0.7pt},
  lbl/.style={font=\scriptsize\itshape, fill=white, inner sep=1pt, text=gray!60},
]

\node[offbox] (data) {%
  \textbf{Data Generation}\par\smallskip
  \footnotesize
  $\cdot$\ Det.UC $\mathbf{u}^*\!\in\!\{0,1\}^{n_g \times T}$\par
  $\cdot$\ Perturbations $\|\mathbf{u}-\mathbf{u}^*\|_1 \leq \epsilon$\par
  $\cdot$\ Scenarios $\Xi = \{\xi_1,\ldots,\xi_K\}$\par
  $\cdot$\ Label $Q(\mathbf{u},\Xi)$ via dispatch%
};

\node[offbox, right=0.9cm of data] (train) {%
  \textbf{ICNN Training}\par\smallskip
  \footnotesize
  $\cdot$\ Scenario embedding\par
  $\cdot$\ ICNN body, convex in $\mathbf{x}$\par
  $\cdot$\ $W_h\!\geq\!0$ (projected gradient)\par
  $\cdot$\ 50-trial hyperparameter search%
};

\node[onbox, right=0.9cm of train] (optim) {%
  \textbf{Neural-Benders Optim.}\par\smallskip
  \footnotesize
  $\cdot$\ First-stage UC + ICNN-MILP\par
  $\cdot$\ Solve $\;\to\;$ candidate $\mathbf{x}$\par
  $\cdot$\ 2nd-stage subproblem + cut\par
  $\cdot$\ Gap certificate $\varepsilon \leq \varepsilon_{\max}$%
};

\draw[arr] (data.east) --
  node[lbl, above=2pt] {$\{(\mathbf{z},\,Q(\mathbf{z}))\}$}
  (train.west);
\draw[arr] (train.east) --
  node[lbl, above=2pt] {$\hat{Q}(\,\cdot\,;\theta^*)$}
  (optim.west);

\node[right=0.7cm of optim, align=left, font=\small] (out)
  {$\mathbf{x}^*$\\[2pt]\footnotesize$\varepsilon \leq \varepsilon_{\max}$};
\draw[arr] (optim.east) -- (out.west);

\draw[-Stealth, draw=blue!50!black, line width=0.6pt]
  ([yshift=-0.40cm]optim.east)
  -- ++(0.32cm,0)
  -- ++(0,-0.55cm)
  -- ++(-0.32cm,0);
\node[right=0.36cm of optim.east, yshift=-0.85cm, anchor=west, align=left,
      font=\scriptsize\itshape, text=blue!55!black] {iterate until\\convergence};

\coordinate (sep) at ($(train.north east)!0.5!(optim.north west)$);
\draw[dashed, gray!45, line width=0.45pt]
  ($(sep)+(0,0.55)$) -- ($(sep)+(0,-0.4)$);

\node[above=0.18cm of data.north, font=\scriptsize\itshape, text=gray!65] {Offline};
\node[above=0.18cm of train.north, font=\scriptsize\itshape, text=gray!65] {Offline};
\node[above=0.18cm of optim.north, font=\scriptsize\itshape, text=blue!55!black] {Online};

\end{tikzpicture}
\caption{Overview of the proposed methodology. The offline phase generates labeled commitment schedules $(\mathbf{u}, Q(\mathbf{u},\Xi))$ and trains the ICNN surrogate. The online phase embeds the trained surrogate into a first-stage MILP and refines the solution via the Neural-Benders loop until the optimality gap $\varepsilon$ is certified.}
\label{fig:methodology}
\end{figure*}

The proposed methodology combines three components: a permutation-invariant scenario aggregation network (Section~\ref{sec:aggregation}), an ICNN surrogate of the recourse cost convex in the first-stage decision by construction (Section~\ref{sec:icnn}), and a Neural-Benders correction loop that certifies solution quality a posteriori (Section~\ref{sec:benders}). An overview is provided in Fig.~\ref{fig:methodology}.

\subsection{Scenario Encoding and Aggregation}
\label{sec:aggregation}
A central challenge in learning surrogates for two-stage stochastic programs is handling a variable number of scenarios without inflating the downstream optimization problem. Following Neur2SP \cite{Dumouchelle2022}, the scenario set is encoded into a compact, fixed-dimensional representation independent of its cardinality.

Each scenario $\xi_k$, $k = 1, \ldots, K$, is independently encoded by a \emph{scenario encoding network} $\Psi_1 : \mathbb{R}^{n_\xi} \to \mathbb{R}^{d_1}$ with shared weights, producing a latent embedding $e_k = \Psi_1(\xi_k)$. The individual embeddings are then pooled into a single aggregate vector via a probability-weighted mean
\begin{equation}
    \label{eq:aggregation}
    a = \sum_{k=1}^{K} p_k \, e_k \;\in\; \mathbb{R}^{d_1},
\end{equation}
where $p_k$ is the probability weight of scenario $k$. Since the dimension of $a$ is independent of $K$, all subsequent computations remain tractable regardless of the scenario set size. An \emph{aggregation network} $\Psi_2 : \mathbb{R}^{d_1} \to \mathbb{R}^{d_2}$ further compresses $a$ into a final scenario representation $z = \Psi_2(a)$. The complete encoding is
\begin{equation}
    \label{eq:encoding}
    z = \Psi_2\!\left( \sum_{k=1}^{K} p_k \, \Psi_1(\xi_k) \right).
\end{equation}
The probability-weighted mean in \eqref{eq:aggregation} is a symmetric operator, guaranteeing permutation invariance over the scenario set. Crucially, since the scenarios $\{\xi_k, p_k\}$ are known parameters at solve time, $\Psi_1$ and $\Psi_2$ are evaluated entirely offline; only the prediction network, which takes $z$ and the first-stage decision $x$ as inputs, requires online computation. We refer the reader to \cite{Dumouchelle2022} for further details on the architecture.

\subsection{Input Convex Neural Networks Architecture}
\label{sec:icnn}

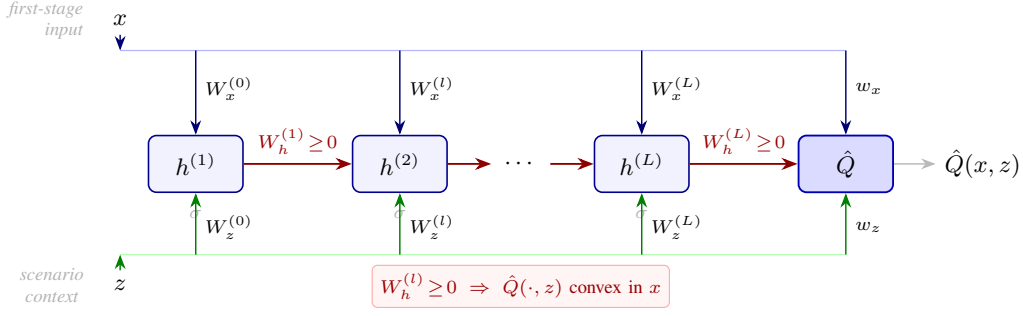
\begin{figure*}[!t]
\centering
\begin{tikzpicture}[
  font=\small, >=Stealth,
  layer/.style={
    draw=blue!55!black, rounded corners=3pt, fill=blue!5,
    minimum width=1.25cm, minimum height=0.75cm,
    inner sep=4pt, align=center, line width=0.6pt,
  },
  outlayer/.style={
    draw=blue!70!black, rounded corners=3pt, fill=blue!14,
    minimum width=1.25cm, minimum height=0.75cm,
    inner sep=4pt, align=center, line width=0.7pt,
  },
  arr/.style={-Stealth, draw=gray!55, line width=0.55pt},
  xarr/.style={-Stealth, draw=blue!45!black, line width=0.55pt},
  zarr/.style={-Stealth, draw=green!50!black, line width=0.55pt},
  wharr/.style={-Stealth, draw=red!55!black, line width=0.70pt},
]

\node[font=\small] (x) at (0.5,  1.9) {$x$};
\node[font=\small] (z) at (0.5, -1.6) {$z$};

\node[layer]    (h1)   at (1.5, 0) {$h^{(1)}$};
\node[layer]    (h2)   at (4.2, 0) {$h^{(2)}$};
\node           (dots) at (5.8, 0) {$\cdots$};
\node[layer]    (hL)   at (7.4, 0) {$h^{(L)}$};
\node[outlayer] (Q)    at (10.1, 0) {$\hat{Q}$};

\node[below=0.10cm of h1,  font=\scriptsize, text=gray!60] {$\sigma$};
\node[below=0.10cm of h2,  font=\scriptsize, text=gray!60] {$\sigma$};
\node[below=0.10cm of hL,  font=\scriptsize, text=gray!60] {$\sigma$};

\draw[blue!30, line width=0.45pt] (0.5, 1.5) -- (10.1, 1.5);
\draw[xarr] (x.south) -- (0.5, 1.5);

\draw[xarr] (1.5, 1.5) -- node[right, font=\scriptsize, pos=0.45]{$W_x^{(0)}$} (h1.north);
\draw[xarr] (4.2, 1.5) -- node[right, font=\scriptsize, pos=0.45]{$W_x^{(l)}$}  (h2.north);
\draw[xarr] (7.4, 1.5) -- node[right, font=\scriptsize, pos=0.45]{$W_x^{(L)}$}  (hL.north);
\draw[xarr] (10.1,1.5) -- node[right, font=\scriptsize, pos=0.45]{$w_x$}          (Q.north);

\draw[green!30, line width=0.45pt] (0.5, -1.2) -- (10.1, -1.2);
\draw[zarr] (z.north) -- (0.5, -1.2);

\draw[zarr] (1.5, -1.2) -- node[right, font=\scriptsize, pos=0.45]{$W_z^{(0)}$} (h1.south);
\draw[zarr] (4.2, -1.2) -- node[right, font=\scriptsize, pos=0.45]{$W_z^{(l)}$}  (h2.south);
\draw[zarr] (7.4, -1.2) -- node[right, font=\scriptsize, pos=0.45]{$W_z^{(L)}$}  (hL.south);
\draw[zarr] (10.1,-1.2) -- node[right, font=\scriptsize, pos=0.45]{$w_z$}          (Q.south);

\draw[wharr] (h1.east) -- node[above, font=\scriptsize, text=red!60!black]{$W_h^{(1)}\!\geq\!0$} (h2.west);
\draw[wharr] (h2.east) -- ($(dots.west)-(0.05,0)$);
\draw[wharr] ($(dots.east)+(0.05,0)$) -- (hL.west);
\draw[wharr] (hL.east) -- node[above, font=\scriptsize, text=red!60!black]{$W_h^{(L)}\!\geq\!0$} (Q.west);

\node[right=0.55cm of Q, font=\small] (output) {$\hat{Q}(x,z)$};
\draw[arr] (Q.east) -- (output);

\node[draw=red!35, rounded corners=2pt, fill=red!4,
      font=\scriptsize, text=red!60!black, inner sep=3pt]
  at (5.8, -1.62)
  {$W_h^{(l)}\!\geq\!0\;\Rightarrow\;\hat{Q}(\cdot,z)$ convex in $x$};

\node[left=0.15cm of x, font=\scriptsize\itshape, text=gray!60, align=right] {first-stage\\input};
\node[left=0.15cm of z, font=\scriptsize\itshape, text=gray!60, align=center] {scenario\\context};

\end{tikzpicture}
\caption{ICNN surrogate architecture (one unrolled path). The first-stage schedule $x$ connects to \emph{every} layer via unconstrained passthrough weights $W_x^{(l)}$ (vertical arrows from the upper bus). The scenario context $z$ similarly connects to \emph{every} layer via weights $W_z^{(l)}$ (vertical arrows from the lower bus). Between consecutive layers, the passthrough weights $W_h^{(l)}$ (horizontal arrows) are constrained non-negative, which, combined with the skip connections from $x$, guarantees convexity of $\hat{Q}$ in $x$ for any fixed $z$.}
\label{fig:icnn_arch}
\end{figure*}

Let $x \in \mathbb{R}^{n_x}$ be the first-stage decision vector and $z \in \mathbb{R}^{d_2}$ the aggregated scenario representation produced offline by $\Psi_2$ (Section~\ref{sec:aggregation}). The ICNN \cite{Amos2017} prediction network, depicted in Fig.~\ref{fig:icnn_arch}, computes
\begin{subequations}
\label{eq:icnn}
\begin{align}
    h^{(1)} &= \sigma\!\left( W_x^{(0)} x + W_z^{(0)} z + b^{(0)} \right), \label{eq:icnn_first} \\
    h^{(l+1)} &= \sigma\!\left( W_h^{(l)} h^{(l)} + W_x^{(l)} x + W_z^{(l)} z + b^{(l)} \right), \label{eq:icnn_hidden} \\
    & \hspace{2em} l = 1,\ldots,L-1, \nonumber
\end{align}
\end{subequations}
where $\sigma(t) = \max\{0,t\}$ is the ReLU activation, the matrices $W_x^{(l)}$, $W_z^{(l)}$ are unconstrained, and the passthrough weight matrices satisfy
\begin{equation}
    W_h^{(l)} \geq 0 \quad \text{(element-wise)}, \quad l = 1,\ldots,L-1. \label{eq:nonneg}
\end{equation}
The output layer produces a scalar prediction
\begin{equation}
    \hat{Q}(x, z) = w_h^\top h^{(L)} + w_x^\top x + w_z^\top z + b, \quad w_h \geq 0. \label{eq:icnn_output}
\end{equation}

The non-negativity of the passthrough weights $W_h^{(l)}$ and output weights $w_h$ is the key architectural constraint. It guarantees that $\hat{Q}(x,z)$ is convex in $x$ for any fixed $z$ \cite{Amos2017}.

This architectural choice is motivated by three structural properties that the standard MLP surrogate of Neur2SP~\cite{Dumouchelle2022} cannot provide. First, in the continuous-recourse setting $Q(x, \xi_s)$ is convex and piecewise linear in $x$ for each fixed scenario $s$ (Section~\ref{sec:twostage}), so the ICNN (differently from MLP) approximates the target within the same function class. Second, convexity by construction eliminates spurious basins from the first-stage MILP that a non-convex surrogate could otherwise introduce. Third, the convex continuous relaxation of the embedded ICNN MILP yields tighter LP bounds per master iteration, reducing master solve times.

In the integer-recourse setting (Section~\ref{sec:strategic}), $Q(x, \xi_s)$ is generally non-convex in $x$. The ICNN provides a convex approximation induced by the architecture and the regression objective, and may both over- and underestimate $Q$ at unseen points. Certification of solution quality is delegated to the Neural-Benders correction loop of Section~\ref{sec:benders}, which provides an a posteriori gap guarantee regardless of the surrogate's local accuracy.

Since the ICNN employs ReLU activations, it admits an exact reformulation as a set of mixed-integer linear constraints \cite{Fischetti2018, Anderson2020}. For each neuron $j$ in layer $l$ with pre-activation $h_j^{(l)}$ and output $z_j^{(l)} = \max\{0, h_j^{(l)}\}$, introducing a binary indicator $\delta_j^{(l)} \in \{0,1\}$ and a large constant $M > 0$, we obtain the system
\begin{subequations}
\label{eq:relu_mip}
\begin{align}
    z_j^{(l)} - h_j^{(l)} &\geq 0, \\
    z_j^{(l)} - h_j^{(l)} &\leq M\, \delta_j^{(l)}, \\
    z_j^{(l)} &\leq M\bigl(1 - \delta_j^{(l)}\bigr).
\end{align}
\end{subequations}
The non-negativity constraints \eqref{eq:nonneg} are enforced during offline training, adding no overhead at the online optimization stage. The resulting surrogate first-stage optimization is
\begin{equation}
    \label{eq:surrogate_mip}
    \min_{x \in X} \left\{ c^\top x + \hat{Q}(x, z) \right\},
\end{equation}
subject to the operational constraints defining $X$, the ICNN forward-pass equations \eqref{eq:icnn} and \eqref{eq:icnn_output}, and the ReLU linearization \eqref{eq:relu_mip}.

\subsection{Neural-Benders Correction Loop}
\label{sec:benders}
The Neural-Benders correction loop is the mechanism designed to recover optimality from an approximate surrogate.

The surrogate optimization \eqref{eq:surrogate_mip} yields a first-stage candidate in substantially reduced time, but the quality of this solution depends on the accuracy of the learned approximation, which may degrade for instances far from the training distribution. A key observation is that approximation errors are not symmetric in their consequences. \emph{Underestimation} ($\hat{Q}(x,z) < \mathbb{E}_\xi[Q(x,\xi)]$) causes the solver to favor regions of artificially low predicted cost; a Benders optimality cut generated at the resulting candidate raises the lower-bound approximation and directly corrects the error. \emph{Overestimation} ($\hat{Q}(x,z) > \mathbb{E}_\xi[Q(x,\xi)]$) has a more insidious effect. The solver perceives the region as artificially expensive, avoids it entirely, generates no corrective information there, and the bias persists undetected. The Neural-Benders loop addresses both failure modes through a penalized slack relaxation coupled with iterative Benders cut generation.

Rather than imposing the ICNN as a hard constraint, we introduce a slack variable $\lambda \geq 0$ that permits the solver to override the neural prediction at a controlled cost. Defining an auxiliary recourse estimator $\eta$ and a penalty parameter $\rho > 0$, the hybrid master problem is
\begin{equation}
    \label{eq:hybrid_mip}
    \min_{x \in X,\, \eta,\, \lambda \geq 0} \left\{ c^\top x + \eta + \rho\,\lambda \right\},
\end{equation}
subject to
\begin{align}
    \eta &\geq \hat{Q}(x, z) - \lambda, \label{eq:hybrid_nn} \\
    \eta &\geq \theta(x), \label{eq:hybrid_cuts}
\end{align}
together with the ICNN MILP constraints \eqref{eq:relu_mip} and the operational constraints defining $X$. Here $\theta(x)$ denotes the pointwise maximum over all accumulated Benders cuts. The penalty $\rho$ governs the trust placed in the learned model. As $\rho \to \infty$, the slack becomes prohibitively expensive ($\lambda^* = 0$) and the ICNN acts as an inviolable lower bound, recovering the hard-constraint surrogate \eqref{eq:surrogate_mip}; as $\rho \to 0$, the ICNN constraint becomes non-binding and \eqref{eq:hybrid_mip}--\eqref{eq:hybrid_cuts} reduces to a classical Benders master problem. For intermediate $\rho$, the solver can enter an overestimated region by paying $\rho$ per unit of relaxation, whenever the improvement in the first-stage cost outweighs this penalty. Once the solver enters such a region, the exact subproblem generates a cut there that reveals the true, lower cost and permanently corrects the bias. The slack never compromises the validity of the lower bound, as it is provided by $\theta(x)$, the accumulated exact-subproblem cuts, not by the ICNN term $\hat{Q}(x,z)-\lambda$; the slack only redirects the search.

The full refinement algorithm structure is given in Algorithm~\ref{alg:neural_benders}; the specific cut forms depend on the recourse setting and are detailed below.

\begin{algorithm}[!t]
\caption{Neural-Benders Correction Loop}
\label{alg:neural_benders}
\begin{algorithmic}[1]
\Require ICNN $\hat{Q}(x,z)$; scenarios $\{\xi_k,p_k\}_{k=1}^{K}$; tolerance $\varepsilon$;
         initial penalty $\rho_0$; no-improvement window $W$; floor $\rho_{\min}$
\Ensure Certified first-stage schedule $x^*$, optimality gap $\mathrm{gap}^*$
\State Initialize $\theta(x)\!\leftarrow\!-\infty$,\; $\mathrm{UB}\!\leftarrow\!+\infty$,\; $\rho\!\leftarrow\!\rho_0$,\; $i\!\leftarrow\!0$
\Repeat
  \State $i \leftarrow i+1$
  \State \textbf{Master solve:} solve \eqref{eq:hybrid_mip}--\eqref{eq:hybrid_cuts} $\;\to\;$ candidate $x^{(i)}$
  \For{$k = 1,\ldots,K$}
    \State \textbf{Subproblem:} solve $Q(x^{(i)},\xi_k)$ exactly (LP or MILP)
  \EndFor
  \State $\overline{Q}(x^{(i)}) \leftarrow \sum_{k=1}^{K} p_k\, Q(x^{(i)},\xi_k)$;\; update $x^{(i)}_{\mathrm{best}}$ and $\mathrm{UB}^{(i)}$
  \State \textbf{Cut:} extract Benders or combinatorial cut from subproblem information; add to $\theta(x)$
  \State $\mathrm{LB}^{(i)} \leftarrow c^{\top}x^{(i)} + \theta(x^{(i)})$;\quad $\mathrm{gap}^{(i)} \leftarrow \mathrm{UB}^{(i)} - \mathrm{LB}^{(i)}$
  \If{no improvement in $\mathrm{UB}$ for $W$ consecutive iterations}
    \State $\rho \leftarrow \max\!\left(\rho/2,\;\rho_{\min}\right)$
  \EndIf
\Until{$\mathrm{gap}^{(i)} \leq \varepsilon$}
\State \Return $x^* \leftarrow x^{(i)}_{\mathrm{best}}$,\; $\mathrm{gap}^* \leftarrow \mathrm{gap}^{(i)}$
\end{algorithmic}
\end{algorithm}

In the continuous-recourse setting, each subproblem is an LP. When the LP is feasible, the optimal dual variables $\pi_k^*$ yield the Benders optimality cut
\begin{equation}
    \label{eq:benders_cut}
    \theta \geq \sum_{k=1}^{K} p_k \left( \pi_k^{*\top} h_k - \pi_k^{*\top} T_k\, x \right),
\end{equation}
a linear inequality valid for all $x \in X$, where $h_k = h(\xi_k)$ and $T_k = T(\xi_k)$. When instead the LP is infeasible at $x^{(i)}$ (as may occur in the absence of complete recourse), the dual unbounded ray $d_k^*$ yields the Farkas feasibility cut
\begin{equation}
    \label{eq:farkas_cut}
    d_k^{*\top}\!\left( h_k - T_k\, x \right) \leq 0,
\end{equation}
added as a hard constraint on $x$ to exclude first-stage decisions that render scenario $k$ infeasible.

In the integer-recourse setting of Section~\ref{sec:strategic}, complete recourse holds since fast-start generators can always decline to activate ($u^f_{g,t,s} = 0$ for all $g \in \mathcal{G}_f$), so Farkas feasibility cuts never arise. The second-stage subproblem at a candidate $x^{(i)}$ is, however, a MILP rather than an LP, so LP duality arguments are no longer applicable and classical Benders optimality cuts lose their validity.

To restore correctness, we adopt the Integer L-Shaped Method of~\cite{Laporte1993}. Before the main loop, a global lower bound $L$ on the recourse function is computed once by relaxing all binary variables, both first-stage ($u_{g,t}$) and second-stage ($u^f_{g,t,s}$), to the unit interval and solving the resulting LP. This bound satisfies $L \leq Q(x, \xi_s)$ for all $x \in X$ and all $s$.

At each iteration, the true recourse cost $Q(x^{(i)}) = \sum_{k=1}^{K} p_k\, Q(x^{(i)}, \xi_k)$ is obtained by solving the full stochastic MILP with $x^{(i)}$ fixed. Defining $S^+ = \{(g,t) : u^{(i)}_{g,t} = 1\}$ and $S^- = \{(g,t) : u^{(i)}_{g,t} = 0\}$, the resulting combinatorial cut is
\begin{equation}
    \label{eq:integer_lshaped_cut}
    \theta \;\geq\; \bigl(Q(x^{(i)}) - L\bigr) \Bigl( \sum_{S^+} u_{g,t} - \sum_{S^-} u_{g,t} - |S^+| + 1 \Bigr) + L.
\end{equation}
This inequality is tight when $x = x^{(i)}$ (Hamming distance zero, inner sum equals one) and relaxes to $\theta \geq L$ for any commitment that differs from $x^{(i)}$ in at least one entry, providing a globally valid lower bound \cite{Laporte1993}. Unlike LP-based cuts, which are hyperplanes in $x$, the Integer L-Shaped cut is combinatorial. It encodes exact recourse information at a specific binary point and decays gracefully elsewhere, at the cost of requiring one MILP solve per iteration rather than one LP solve.

Each iteration solves all $K$ second-stage subproblems, exactly as classical Benders and Integer L-Shaped decomposition do, so the per-iteration cost is identical; the advantage lies in the iteration count. Since the ICNN warm start has no spurious basins, the initial candidate is already close to the optimum and the loop certifies convergence in 1--13 iterations across the benchmark (Tables~\ref{tab:results_continuous}--\ref{tab:results_integer}), against up to 100 uncertified iterations for the classical Integer L-Shaped method, i.e., far fewer total subproblem solves.

The scenario evaluation also provides a computable upper bound on the true expected recourse cost at the candidate, i.e.
\begin{equation}
    \label{eq:upper_bound}
    \overline{Q}(x^{(i)}) = \sum_{k=1}^{K} p_k\, Q\!\left(x^{(i)}, \xi_k\right).
\end{equation}
Letting $x^{(i)}_{\mathrm{best}}$ denote the best incumbent found up to iteration $i$, the solution quality bounds and gap are
\begin{align}
    \mathrm{UB}^{(i)} &= c^\top x^{(i)}_{\mathrm{best}} + \overline{Q}(x^{(i)}_{\mathrm{best}}), \label{eq:ub_def} \\
    \mathrm{LB}^{(i)} &= c^\top x^{(i)} + \theta(x^{(i)}), \label{eq:lb_def} \\
    \mathrm{gap}^{(i)} &= \mathrm{UB}^{(i)} - \mathrm{LB}^{(i)}, \label{eq:quality_gap}
\end{align}
where $\theta(x^{(i)})$ is the pointwise maximum of all accumulated cuts at the current master iterate $x^{(i)}$. The certified lower bound $\mathrm{LB}^{(i)}$ is taken from the cut-only relaxation of the master, not from the penalized hybrid objective, which carries the $\rho\lambda$ and $\hat{Q}$ terms and is not a clean bound on the true recourse cost. The iteration terminates when $\mathrm{gap}^{(i)} \leq \varepsilon$. When convergence stagnates, signaling persistent overestimation, $\rho$ is reduced adaptively, progressively ceding control to the Benders cuts and recovering the behavior of classical Benders decomposition as $\rho \to 0$. Concretely, $\rho$ is initialised at $\rho_0 = 1.01$, halved after every $W = 3$ consecutive non-improving iterations subject to a floor $\rho_{\min} = 10^{-3}$, and convergence is declared at $\varepsilon = 10^{-3}$ relative gap.

\section{Case Studies}
\label{sec:experiments}
The proposed methodology is evaluated on the two problem classes of Sections~\ref{sec:suc} and~\ref{sec:strategic}, instantiated on the IEEE transmission test systems (9 to 118 buses)~\cite{Zimmermann2011} with $K \in \{10, 50, 100\}$ scenarios, a controlled and reproducible benchmark for assessing methodological soundness and scalability.

\subsection{Stochastic UC on IEEE Transmission Test Systems}
\label{sec:case_ieee}
The first class of experiments is based on the IEEE transmission network test cases, specifically the 9-, 14-, 30-, 39-, 57-, and 118-bus systems \cite{Zimmermann2011}. These benchmarks were originally introduced as Optimal Power Flow (OPF) instances; we use them here in their SUC variant, following the two-stage formulation detailed in Section~\ref{sec:suc}. Generator parameters (capacities, marginal costs, ramp rates, start-up and no-load costs, minimum up/down times) and network data (line susceptances and thermal limits) are taken directly from the MATPOWER data files \cite{Zimmermann2011} without modification.

The uncertain parameter $\xi_s$ is the vector of net loads $d_{b,t,s}$ at each bus $b$ and period $t$. Scenarios are generated by applying a per-period multiplier $\alpha_{t,s} \sim \mathcal{U}[0.7, 1.0]$, drawn independently for each $(t,s)$ and applied uniformly across all buses: $d_{b,t,s} = \alpha_{t,s} \cdot d_b^{\text{base}} \cdot \ell_t$, where $d_b^{\text{base}}$ is the nominal MATPOWER bus load and $\ell_t$ is a 24-hour load shape with morning and evening peaks. Using a single multiplier per time period produces spatially correlated scenarios consistent with the correlated forecast errors typical of day-ahead UC.

\subsection{IEEE Transmission Benchmarks with Integer Recourse}
\label{sec:case_ieee_int}
The second class of experiments applies the integer recourse formulation of Section~\ref{sec:strategic} to the same systems, parameters, and scenarios; the only modification is the slow/fast partition of $\mathcal{G}$, with $p_f = 0.8$ so that the 20\% most expensive generators are fast-start, their commitment being deferred to the second stage as binary recourse variables $u^f_{g,t,s}$.

\subsection{Data Generation}
\label{sec:datagen}

Training the neural surrogate requires a labeled dataset of input-output pairs $\{(x^{(i)},\, z^{(i)},\, \hat{Q}^{(i)})\}$, where $x^{(i)}$ is a first-stage decision, $z^{(i)}$ encodes the stochastic context (sampled scenarios and associated probabilities), and the label $\hat{Q}^{(i)} \approx \mathbb{E}_\xi[Q(x^{(i)},\xi)]$ is obtained by evaluating the recourse cost over a finite scenario bank. All architectures are trained on the same dataset; only the training and surrogate-encoding steps differ.

\noindent\textbf{IEEE Transmission Benchmarks.}
The network maps $(x,\, z)$ to $\hat{Q}$, where $x \in \{0,1\}^{n_g \times T}$ is the commitment schedule and $z \in \mathbb{R}^{d_2}$ is the pre-computed scenario embedding (Section~\ref{sec:methodology}). The label $\hat{Q}^{(i)}$ is the expected production cost under a fixed bank of $K$ load scenarios. Generating diverse, feasible, and informative commitment samples is non-trivial for binary decision spaces; purely random draws rarely satisfy temporal coupling constraints, while optimal solutions alone provide insufficient coverage. We therefore adopt the methodology of~\cite{Shao2025}, which constructs a set of feasible schedules around reference UC solutions through a structured perturbation scheme. Infeasible samples are retained and consistently labeled via dual information extracted from the reference solutions rather than discarded; we refer the reader to~\cite{Shao2025} for full details.

\noindent\textbf{IEEE Benchmarks with Integer Recourse.}
The input-output structure is identical to the continuous case, with $x \in \{0,1\}^{|\mathcal{G}_s| \times T}$ restricted to slow-generator commitments. The same hot-start scheme of~\cite{Shao2025} is applied; the only difference is that each labeling solve is a MILP rather than an LP, since the second stage includes the fast-generator binary variables $u^f_{g,t,s}$.

\subsection{Training Setup}
\label{sec:training}

Both Neur2SP and the proposed ICNN surrogate are trained with the same protocol, minimizing the mean squared error (MSE) on normalized targets. Hyperparameters are selected via parallel random search over 50 trials, with model selection based on the lowest denormalized mean absolute error (MAE) on the validation set. Table~\ref{tab:training_setup} summarises the common training configuration.

\begin{table}[!t]
\renewcommand{\arraystretch}{1.15}
\caption{Training configuration shared by Neur2SP and ICNN.}
\label{tab:training_setup}
\centering
\small
\begin{tabular}{ll}
\toprule
\textbf{Setting} & \textbf{Value} \\
\midrule
Optimizer          & Adam \\
Loss function      & MSE (normalized targets) \\
Train / val split  & 80\,\% / 20\,\% \\
Early stopping     & patience = 50 epochs (val.\ loss) \\
Model selection    & min.\ denormalized val.\ MAE \\
Hyperparameter search & parallel random search, 50 trials \\
\midrule
\multicolumn{2}{l}{\textit{Search space}} \\
\quad Hidden layers        & 2--4 \\
\quad Hidden width         & 32--256 \\
\quad Scenario embed.\ dim & 16--128 \\
\quad Learning rate        & $[10^{-4},\,10^{-2}]$ (log-uniform) \\
\quad Dropout              & $[0.0,\,0.3]$ \\
\quad Batch size           & 8--64 \\
\midrule
\multicolumn{2}{l}{\textit{ICNN-specific}} \\
\quad $W_h$ init.         & $\mathcal{U}(0.01,\,0.1)$ (strictly positive) \\
\quad Convexity enforcement & projected gradient after each step: \\
                            & $W_h \leftarrow \max(W_h,\,0)$ \\
\bottomrule
\end{tabular}
\end{table}

The recourse cost target $\hat{Q}$ is standardized to zero mean and unit variance using a per-case StandardScaler fit on the training set; the same scaler is applied at inference time to recover physical costs. The commitment input $x \in \{0,1\}^{n_g \times T}$ requires no scaling. Offline computational cost (data labeling and hyperparameter search) ranged from a few hours for small cases (\texttt{case9}--\texttt{case14}) to approximately 48 hours for the largest (\texttt{case118}) on the benchmark machine described in Section~\ref{sec:results}. All optimization in the Neural-Benders loop is solved with Gurobi 12.0.

\subsection{Methodology Comparison}
\label{sec:comparison}

Five methods are evaluated in each experimental setting.

\noindent\textbf{Extensive Form (EF).} The reference deterministic-equivalent MILP (Section~\ref{sec:suc}), solved directly. Optimal but computationally the most expensive.

\noindent\textbf{L-Shaped / Integer L-Shaped (LS / i-LS).} The classical Benders decomposition~\cite{Papavasiliou2015} applied to the continuous and integer recourse settings respectively. Both are implemented with a maximum of 100 iterations; cases that do not converge within this limit are marked with $\dagger$ in the tables.

\noindent\textbf{Neur2SP (N2SP).} The baseline surrogate approach of~\cite{Dumouchelle2022}, which embeds a standard ReLU multi-layer perceptron as the recourse cost approximation, trained on the same dataset and encoded as a MILP via big-M constraints.

\noindent\textbf{ICNN Surrogate (ICNN).} The proposed ICNN-based surrogate of Section~\ref{sec:icnn}, without iterative refinement.

\noindent\textbf{Neural-Benders (ICNN+).} The full proposed methodology of Section~\ref{sec:benders}, with a posteriori optimality certificate.


\section{Numerical Results and Discussion}
\label{sec:results}

\begin{table*}[!t]
\renewcommand{\arraystretch}{1.05}
\caption{Continuous recourse. $t$: wall clock time [s]; gap: \% gap vs.\ EF; $\times$: $t_{\mathrm{EF}}/t_{\mathrm{method}}$; \# iter: Benders iterations (LS and ICNN+ only). $\dagger$: hit iteration limit without convergence;  n/c: not converged.}
\label{tab:results_continuous}
\centering
\setlength{\tabcolsep}{2pt}
\begin{tabular}{ll rrrrr rrrr rrrr rr}
\toprule
& & \multicolumn{5}{c}{$t$ [s]} & \multicolumn{4}{c}{gap [\%]} & \multicolumn{4}{c}{$\times$} & \multicolumn{2}{c}{\# iter} \\
\cmidrule(lr){3-7}\cmidrule(lr){8-11}\cmidrule(lr){12-15}\cmidrule(lr){16-17}
Case & $K$ & EF & LS & N2SP & ICNN & ICNN+ & LS & N2SP & ICNN & ICNN+ & LS & N2SP & ICNN & ICNN+ & LS & ICNN+ \\
\midrule
\multirow{3}{*}{case9}
  & 10  &    0.67 &   1.16 & \textbf{0.61} & 0.71 &  3.30 & 0.00 & 0.11 & 0.09 & \textbf{0.00} & 0.58 & \textbf{1.10} &   0.94 &  0.20 & \textbf{$\sim$2} & $\sim$4 \\
  & 50  &    2.96 &   5.83 & \textbf{0.52} & 0.72 &  1.82 & 0.00 & 0.01 & 0.02 & \textbf{0.00} & 0.51 & \textbf{5.69} &   4.11 &  1.63 & \textbf{$\sim$2} & \textbf{$\sim$2} \\
  & 100 &    5.69 &  12.03 & \textbf{0.51} & 0.94 &  2.44 & 0.00 & 0.01 & 0.05 & \textbf{0.00} & 0.47 & \textbf{11.16} &   6.05 &  2.33 & \textbf{$\sim$2} & \textbf{$\sim$2} \\
\midrule
\multirow{3}{*}{case14}
  & 10  &    2.03 &  18.70 & 1.00 & \textbf{0.90} &  2.77 & 0.00 & 0.07 & 0.25 & \textbf{0.00} & 0.11 &   2.03 & \textbf{2.26} &  0.73 & $\sim$26 & \textbf{$\sim$2} \\
  & 50  &    9.10 & 107.97 & \textbf{0.71} & 1.83 &  3.03 & 0.00 & 0.07 & 0.67 & \textbf{0.00} & 0.08 & \textbf{12.82} &   4.97 &  3.00 & $\sim$25 & \textbf{$\sim$2} \\
  & 100 &   17.68 & 227.87 & \textbf{0.97} & 1.01 &  3.80 & 0.00 & 0.14 & 0.77 & \textbf{0.00} & 0.08 & \textbf{18.23} &  17.50 &  4.65 & $\sim$25 & \textbf{$\sim$2} \\
\midrule
\multirow{3}{*}{case30}
  & 10  &    7.02 &  53.94 & \textbf{1.01} & 1.10 &  2.24 & 0.00 & 12.54 & 0.91 & \textbf{0.00} & 0.13 & \textbf{6.95} &   6.38 &  3.13 & $\sim$36 & \textbf{$\sim$6} \\
  & 50  &   33.85 & 642.39 &  0.96 & \textbf{0.79} &  9.66 & 0.00 & 13.56 & 2.76 & \textbf{0.00} & 0.05 &  35.26 & \textbf{42.85} &  3.50 & $\sim$80 & \textbf{$\sim$11} \\
  & 100 &   67.47 & 781.48 &  1.65 & \textbf{1.19} &  15.49 & 0.00 &  8.61 & 4.14 & \textbf{0.00} & 0.09 &  40.89 & \textbf{56.70} & 4.36 & $\sim$49 & \textbf{$\sim$13} \\
\midrule
\multirow{3}{*}{case39}
  & 10  &    8.50 & 178.52$^\dagger$ & \textbf{0.60} & 0.69 &  1.94 & n/c & 0.01 & 0.01 & \textbf{0.00} & 0.05$^\dagger$ & \textbf{14.17} &  12.32 &  4.38 & 100$^\dagger$ & \textbf{$\sim$2} \\
  & 50  &   54.88 & 952.40$^\dagger$ & \textbf{0.54} & 0.74 &  2.64 & n/c & 0.01 & 0.01 & \textbf{0.00} & 0.06$^\dagger$ & \textbf{101.63} &  74.16 & 20.79 & 100$^\dagger$ & \textbf{$\sim$2} \\
  & 100 &  101.11 & 277.40 & \textbf{0.56} & 1.89 &  3.78 & 0.00 & 0.01 & 0.01 & \textbf{0.00} & 0.36 & \textbf{180.55} &  53.50 & 26.75 & $\sim$27 & \textbf{$\sim$2} \\
\midrule
\multirow{3}{*}{case57}
  & 10  &   19.91 &  78.94 &  1.28 & \textbf{0.88} &  3.11 & 0.00 & 0.02 & 0.23 & \textbf{0.00} & 0.25 &  15.55 & \textbf{22.63} &  6.40 & $\sim$28 & \textbf{$\sim$3} \\
  & 50  &  125.61 & 434.51 &  2.46 & \textbf{0.71} &  3.71 & 0.00 & 0.13 & 0.01 & \textbf{0.00} & 0.29 &  51.06 & \textbf{176.92} & 33.86 & $\sim$26 & \textbf{$\sim$2} \\
  & 100 &  265.20 & 739.23 & \textbf{0.59} & 0.81 &  3.75 & 0.00 & 0.01 & 0.01 & \textbf{0.00} & 0.36 & \textbf{449.49} & 327.41 & 70.72 & $\sim$26 & \textbf{$\sim$1} \\
\midrule
\multirow{3}{*}{case118}
  & 10  &  150.53 & 139.08 & \textbf{2.99} &  3.90 &  8.15 & 0.00 & 2.97 & 0.08 & \textbf{0.00} & 1.08 & \textbf{50.34} &  38.60 & 18.47 & $\sim$26 & \textbf{$\sim$2} \\
  & 50  &  843.01 & 181.27 & \textbf{3.10} &  4.04 &  8.66 & 0.00 & 0.01 & 0.08 & \textbf{0.00} & 4.65 & \textbf{271.93} & 208.67 & 97.35 & $\sim$26 & \textbf{$\sim$2} \\
  & 100 & 3018.61 & 411.09 & 2.99 &  \textbf{1.96} &  5.27 & 0.00 & 7.80 & 0.08 & \textbf{0.00} & 7.34 & 1009.57 & \textbf{1540.11} & 572.79 & $\sim$26 & \textbf{$\sim$2} \\
\bottomrule
\end{tabular}
\end{table*}

\begin{table*}[!t]
\renewcommand{\arraystretch}{1.05}
\caption{Integer recourse. $t$: wall clock time [s]; gap: \% gap vs.\ EF; $\times$: $t_{\mathrm{EF}}/t_{\mathrm{method}}$; \# iter: Benders iterations (i-LS and ICNN+ only). $\dagger$: hit iteration limit without convergence; n/c: not converged.}
\label{tab:results_integer}
\centering
\setlength{\tabcolsep}{2pt}
\begin{tabular}{ll rrrrr rrrr rrrr rr}
\toprule
& & \multicolumn{5}{c}{$t$ [s]} & \multicolumn{4}{c}{gap [\%]} & \multicolumn{4}{c}{$\times$} & \multicolumn{2}{c}{\# iter} \\
\cmidrule(lr){3-7}\cmidrule(lr){8-11}\cmidrule(lr){12-15}\cmidrule(lr){16-17}
Case & $K$ & EF & i-LS & N2SP & ICNN & ICNN+ & i-LS & N2SP & ICNN & ICNN+ & i-LS & N2SP & ICNN & ICNN+ & i-LS & ICNN+ \\
\midrule
\multirow{3}{*}{case9}
  & 10  &   0.89 &  33.92$^\dagger$ & \textbf{0.54} & 0.92 & 2.05 & 0.00 & 0.06 & 0.01 & \textbf{0.00} & 0.03$^\dagger$ & \textbf{1.65} & 0.97 & 0.43 & 100$^\dagger$ & \textbf{$\sim$1} \\
  & 50  &   3.03 & 137.69$^\dagger$ & 0.63          & \textbf{0.61} & 2.66 & 8.98 & 0.02 & 0.01 & \textbf{0.00} & 0.02$^\dagger$ &  4.81 & \textbf{4.97} & 1.14 & 100$^\dagger$ & \textbf{$\sim$1} \\
  & 100 &   5.83 & 280.21$^\dagger$ & \textbf{0.51} & 1.27 & 5.45 & 16.65 & 2.16 & 2.17 & \textbf{0.00} & 0.02$^\dagger$ & \textbf{11.43} & 4.59 & 1.07 & 100$^\dagger$ & \textbf{$\sim$1} \\
\midrule
\multirow{3}{*}{case14}
  & 10  &   2.05 &  55.74$^\dagger$ & 1.13  & \textbf{0.69} &  3.89 & n/c & 0.95 & 2.01 & \textbf{0.00} & 0.04$^\dagger$ &  1.81 & \textbf{2.97} & 0.53 & 100$^\dagger$ & \textbf{$\sim$2} \\
  & 50  &   9.15 & 247.34$^\dagger$ & \textbf{1.76} & 1.91 &  4.39 & n/c & 0.98 & 0.43 & \textbf{0.00} & 0.04$^\dagger$ & \textbf{5.20} & 4.79 & 2.08 & 100$^\dagger$ & \textbf{$\sim$1} \\
  & 100 &  18.29 & 509.44$^\dagger$ & 1.25 & \textbf{0.79} & 4.58 & n/c & 1.00 & 0.12 & \textbf{0.00} & 0.04$^\dagger$ & 14.63 & \textbf{23.15} & 3.99 & 100$^\dagger$ & \textbf{$\sim$1} \\
\midrule
\multirow{3}{*}{case30}
  & 10  &   7.10 &  93.58$^\dagger$ & 4.36 & \textbf{1.00} &  4.13 & n/c & 4.15 & 1.20 & \textbf{0.00} & 0.08$^\dagger$ &  1.63 & \textbf{7.10} & 1.72 & 100$^\dagger$ & \textbf{$\sim$2} \\
  & 50  &  40.79 & 451.22$^\dagger$ & 4.84 & \textbf{1.17} &  9.36 & n/c & 5.85 & 3.69 & \textbf{0.00} & 0.09$^\dagger$ &  8.43 & \textbf{34.86} & 4.36 & 100$^\dagger$ & \textbf{$\sim$4} \\
  & 100 &  87.20 & 961.92$^\dagger$ & 5.02 & \textbf{1.28} &  7.56 & n/c & 5.36 & 0.89 & \textbf{0.00} & 0.09$^\dagger$ & 17.37 & \textbf{68.13} & 11.53 & 100$^\dagger$ & \textbf{$\sim$3} \\
\midrule
\multirow{3}{*}{case39}
  & 10  &  13.38 & 117.81$^\dagger$ & \textbf{3.06} & 3.69 &   4.94 & n/c & 0.55 & 2.17 & \textbf{0.00} & 0.11$^\dagger$ & \textbf{4.37} &  3.63 & 2.71 & 100$^\dagger$ & \textbf{$\sim$2} \\
  & 50  &  61.27 & 563.58$^\dagger$ & \textbf{2.45} & 3.74 &  10.64 & n/c & 0.39 & 1.57 & \textbf{0.00} & 0.11$^\dagger$ & \textbf{25.01} & 16.38 & 5.76 & 100$^\dagger$ & \textbf{$\sim$1} \\
  & 100 & 118.56 & 1171.62$^\dagger$ & \textbf{2.53} & 4.89 &  14.78 & n/c & 0.61 & 1.89 & \textbf{0.00} & 0.10$^\dagger$ & \textbf{46.86} & 24.25 & 8.02 & 100$^\dagger$ & \textbf{$\sim$2} \\
\midrule
\multirow{3}{*}{case57}
  & 10  &  27.52 & 195.04$^\dagger$ &  1.68 & \textbf{1.54} &  8.37 & n/c & 1.53 & 2.50 & \textbf{0.00} & 0.14$^\dagger$ & 16.38 & \textbf{17.87} &  3.29 & 100$^\dagger$ & \textbf{$\sim$3} \\
  & 50  & 124.80 & 903.90$^\dagger$ & \textbf{1.49} & 2.07 & 12.37 & n/c & 1.07 & 2.09 & \textbf{0.00} & 0.14$^\dagger$ & \textbf{83.76} & 60.29 & 10.09 & 100$^\dagger$ & \textbf{$\sim$4} \\
  & 100 & 278.43 & 1954.79$^\dagger$ & \textbf{3.52} & 5.71 & 15.31 & n/c & 0.84 & 2.70 & \textbf{0.00} & 0.14$^\dagger$ & \textbf{79.10} & 48.76 & 18.19 & 100$^\dagger$ & \textbf{$\sim$3} \\
\midrule
\multirow{3}{*}{case118}
  & 10  &  180.48 & 510.05$^\dagger$ & \textbf{3.25} & 3.55 &  8.42 & n/c & 0.04 & 0.24 & \textbf{0.00} & 0.35$^\dagger$ &  \textbf{55.53} &  50.84 &  21.43 & 100$^\dagger$ & \textbf{$\sim$1} \\
  & 50  &  969.87 & 2478.46$^\dagger$ & \textbf{2.16} & 3.12 & 12.96 & n/c & 0.02 & 0.23 & \textbf{0.00} & 0.39$^\dagger$ & \textbf{449.01} & 310.86 &  74.84 & 100$^\dagger$ & \textbf{$\sim$1} \\
  & 100 & 3876.20 & 3542.72$^\dagger$ & \textbf{2.47} & 4.75 & 18.15 & n/c & 0.01 & 0.24 & \textbf{0.00} & 1.09$^\dagger$ & \textbf{1569.31} & 815.94 & 213.61 & 100$^\dagger$ & \textbf{$\sim$1} \\
\bottomrule
\end{tabular}
\end{table*}

All experiments were run on an Apple M1 (8-core CPU, 16\,GB RAM); models are implemented in Pyomo and solved with Gurobi~12.0. Each reported figure is the average of 30 runs with different random seeds for scenario sampling.

\subsection{Neural Approximation Quality and Surrogate Performance}
\label{sec:discussion}

On the validation set, pooled across $K \in \{10, 50, 100\}$, both architectures achieve median relative absolute prediction errors below 0.2\% on all cases except \texttt{case30}, where the median error rises to 6.0\% (ICNN) and 4.7\% (Neur2SP) under continuous recourse and to 0.8\% and 0.7\% under integer recourse; this directly explains the larger optimality gaps observed there in Tables~\ref{tab:results_continuous}--\ref{tab:results_integer}. The difficulty is not attributable to problem size alone. Two structural factors likely contribute: the recourse value function in \texttt{case30} may be less regular, with abrupt changes in active constraint sets as the commitment vector varies, making it harder to approximate with a compact surrogate; and the topology may exhibit a higher density of near-equivalent commitment configurations with substantially different redispatch costs, creating sharp decision boundaries that are difficult to capture globally. Crucially, both surrogates are affected, which points to a property of the function being learned rather than to a limitation of either model class specifically. The computational consequences of this harder approximation are discussed in Section~\ref{sec:discussion_comp}. Excluding \texttt{case30}, ICNN and Neur2SP achieve comparable errors, indicating that the convexity constraint restricting the function class costs little in accuracy while providing class consistency and a spurious-basin-free optimization landscape.

\subsection{Computational Performance and Solution Quality}
\label{sec:discussion_comp}

Tables~\ref{tab:results_continuous} and~\ref{tab:results_integer} report wall clock time $t$, average percentage gap with respect to the EF reference, speedup $\times = t_{\mathrm{EF}}/t_{\mathrm{method}}$, and the number of Benders iterations (\# iter) required by ICNN+ to reach a certified zero gap, for all IEEE test cases and scenario dimensions $K \in \{10, 50, 100\}$.

Neur2SP and ICNN consistently deliver sub-second solve times across almost all instances and scenario dimensions, achieving speedups up to $450\times$ in the continuous recourse setting (\texttt{case57}, $K=100$) and up to $1570\times$ in the integer recourse setting (\texttt{case118}, $K=100$). On well-approximated instances, both methods maintain a percentage gap below 1\% with respect to EF. ICNN tends to exhibit slightly higher solve times than Neur2SP. The convexity constraint imposed on the network architecture typically requires deeper or wider networks to achieve comparable approximation accuracy, which translates into a larger MILP embedding at solve time. The same constraint also restricts the representational capacity of ICNN relative to Neur2SP, resulting in somewhat higher optimality gaps on certain instances. These gaps are, however, systematically corrected to zero by the Neural-Benders refinement in ICNN+. ICNN+ is therefore inherently slower than ICNN, since its total solve time comprises the ICNN solve plus the refinement loop; yet the refinement adds only marginal computational overhead (typically between 1 and 13 iterations) while providing the critical advantage of certifying a zero optimality gap a posteriori. This makes ICNN+ the only method in the comparison capable of guaranteeing solution quality without relying on surrogate accuracy.

A further advantage of the surrogate-based methods is their computational stability. EF exhibited substantial variability in solve time across scenario realizations; for \texttt{case118} at $K=100$, individual runs ranged from approximately 2\,300\,s to 4\,300\,s in the integer setting, with the reported values being averages over 30 runs. The surrogate methods, by contrast, produced stable solve times across all random seeds, a property that is practically valuable in operational contexts where predictable latency matters.

The solve-time advantage of surrogate methods is contingent on controlling the complexity of the trained network. Computational performance at inference time depends primarily on the architecture selected during hyperparameter search; deeper networks with more neurons per layer improve approximation quality but produce larger MILP embeddings with more binary variables, potentially eroding the speedup advantage over EF. Although the surrogates show stable performance as the test case size and number of scenarios increase, approximating more complex recourse functions will likely require larger architectures, making it necessary to bound the search space during training. In this context, ICNN+ offers a particularly attractive design trade-off; by keeping the ICNN architecture small and delegating solution improvement to the Neural-Benders refinement loop, compact MILP embeddings are maintained while guaranteeing certified zero-gap solutions, a structural advantage unavailable to uncertified surrogate methods.

\texttt{case30} is the hardest instance (Section~\ref{sec:discussion}); both surrogates face a hard approximation problem on this topology, leading to double-digit gaps for Neur2SP and multi-percent gaps for ICNN. ICNN+ corrects the solution in every instance but at the cost of more Benders iterations (up to $\sim$13 for $K=100$ in the continuous setting), confirming that convergence speed is sensitive to surrogate quality. The certified zero gap achieved by ICNN+ on \texttt{case30}, despite a median surrogate error of up to 6\% (Section~\ref{sec:discussion}), confirms that the quality certificate holds regardless of surrogate accuracy.

A key advantage of ICNN+ over LS and i-LS is its ability to reach certified convergence within a small number of iterations across all tested instances. In contrast, LS fails to converge in several continuous-recourse cases (e.g., \texttt{case39} at $K \leq 50$), and i-LS consistently hits the 100-iteration limit without convergence across the entire integer-recourse benchmark. The surrogate initialization provides a warm start that substantially reduces the number of Benders cuts required, making convergence tractable where the classical decomposition alone is not.

From a practical standpoint, the solve times of ICNN+, while slightly higher than those of the pure surrogate methods, remain compatible with day-ahead time windows: even in the most demanding instance (\texttt{case118}, $K=100$), ICNN+ completes in under 20\,s with a certified zero optimality gap on the SAA instance.

\section{Conclusion}
\label{sec:conclusion}

We proposed a surrogate-based framework for two-stage stochastic UC that combines an ICNN surrogate with a Neural-Benders refinement loop to deliver certified solutions. The ICNN enforces convexity of the recourse cost in the first-stage decision, enabling MILP embedding and exact Benders cuts that certify optimality a posteriori, and extends the approach to integer recourse where classical decompositions fail.

Experiments on the IEEE benchmark suite (\texttt{case9}--\texttt{case118}), under both continuous and integer recourse, confirm certified zero optimality gap on the $K$-scenario SAA instance across all test cases and scenario dimensions, with speedups up to $214\times$ over EF in the integer setting. Solve times are stable across scenario realizations and compatible with day-ahead time windows on the tested benchmark instances. Future work could extend this methodology to multi-stage stochastic problems in power systems.

\section*{Acknowledgment}
The authors acknowledge the use of an AI-based writing assistant to refine the wording and clarity of the text in all sections of this manuscript. The methodology, experiments, analysis, and conclusions are entirely the authors' own, and all AI-assisted edits were reviewed and approved by the authors.


%

\ifCLASSOPTIONcaptionsoff
  \newpage
\fi



%

%

\end{document}